\documentclass[pre,twocolumn,superscriptaddress]{revtex4}
\usepackage{graphicx}
\usepackage{graphics}
\usepackage{latexsym}
\usepackage{amsmath, amsthm, amssymb, wasysym}
\usepackage{hyperref,xcolor}
\usepackage{epsfig}
\usepackage[mathscr]{euscript}
\usepackage{blindtext}
\usepackage[normalem]{ulem}

\begin{document}

\title{Escape of a lamb to safe haven in pursuit by a lion under restarts}

\author{R. K. Singh}
\email{rksinghmp@gmail.com}
\affiliation{Department of Physics, Bar-Ilan University, Ramat-Gan 5290002, Israel}

\author{T. Sandev}
\email{trifce.sandev@manu.edu.mk}
\affiliation{Research Center for Computer Science and Information Technologies,
Macedonian Academy of Sciences and Arts, Bul. Krste Misirkov 2, 1000 Skopje, Macedonia}
\affiliation{Institute of Physics \& Astronomy, University of Potsdam, D-14776 Potsdam-Golm, Germany}
\affiliation{Institute of Physics, Faculty of Natural Sciences and Mathematics,
Ss.~Cyril and Methodius University, Arhimedova 3, 1000 Skopje, Macedonia}

\author{Sadhana Singh}
\email{sdhnsingh080@gmail.com}
\affiliation{The Avram and Stella Goldstein-Goren Department of Biotechnology Engineering,
Ben-Gurion University of the Negev, Be’er Sheva 84105, Israel}

\begin{abstract}
We study the escape behavior of a lamb to safe haven pursued by a hungry
lion. Identifying the system with a pair of vicious Brownian walkers we evaluate the
probability density function for the vicious pair and from there we estimate the
distribution of first passage times. The process ends in two ways: either the lamb
makes it to the safe haven (success) or is captured by the lion (failure). We find that
the conditional distribution for both success and failure possesses a finite mean,
but no higher moments exist. This makes it interesting to study these first passage
properties of this Bernoulli process under restarts, which we do via Poissonian and
sharp restart protocols. We find that under both restart protocols the probability
of success exhibits a monotonic dependence on the restart parameters, however, their
approach to the case without restarts is completely different. The distribution
of first passage times exhibits an exponential decay for the two restart protocols.
In addition, the distribution under sharp resetting also exhibits a periodic behavior,
following the periodicity of the sharp restart protocol itself.
\end{abstract}

\maketitle

\bibliographystyle{apsrev4-1}

\newcommand{\dd}{\partial}
\newcommand{\fr}{\frac}
\newcommand{\tl}{\tilde}

\textit{Introduction:}
Capture processes are one of the classic problems studied within the realm of nonequilibrium
statistical physics \cite{krapivsky2010kinetic,bernardi2022run} with applications ranging from reaction
systems \cite{elgart2004rare,assaf2006spectral,assaf2007spectral} to population dynamics
\cite{khasin2009extinction} to kinetochore capture by spindle molecules \cite{nayak2020comparison}.
As vicious walkers destroy each other the moment their paths cross,
they provide a natural setting to study the properties of capture processes
\cite{fisher1984walks,huse1984commensurate,ispolatov1996war,bray2013persistence,forrester1989vicious,
katori2002scaling,baik2000random,essam1995vicious,pedersen2009bubble}.
The viciousness property of two particles can be described as the chemical reaction:
$A + A \to \phi$ \cite{cardy1996theory,fisher1988reunions,schehr2008exact,kundu2014maximal,
cardy2003families} and has been applied to study the classic lion-lamb capture problem in
which a hungry lion pursues a diffusive prey \cite{krapivsky1996kinetics,redner1999capture,
nayak2020capture}. The quantity of primary interest in the realm of capture problems
is the survival probability of the the evasive prey \cite{oshanin2009survival} and is often
estimated via the method of images employed in a wide class of first passage problems
\cite{redner2001guide} including capture problems \cite{redner1999capture,redner2014gradual,
ben2003ordering}. While it is certain in one dimension that the prey will eventually be killed,
owing to recurrence \cite{klafter2011first,weiss1983random}, a safe haven can provide
a life saving opportunity \cite{gabel2012can}.

Even though the escape of a lamb to a safe haven in pursuit by a hungry lion is of interest
in its own right, the stochastic process itself belongs to the broad category of Bernoulli
trials \cite{feller1971introduction}. A stochastic process is termed as a Bernoulli trial
if it can end in two ways. Examples including but not limited to are
the gambler ruin problem \cite{edwards1983pascal},
multiple targets in confined geometries \cite{condamin2007random,condamin2008probing,
benichou2015mean}, mortal random walkers \cite{lohmar2009diffusion,abad2012survival,
yuste2013exploration,campos2015optimal,meerson2015mortality}, chemical selectivity
\cite{rehbein2011we}, multiple folding options for a biopolymer \cite{solomatin2010multiple,
pierse2017distinguishing}. It is not unusual to designate a desired outcome of such
stochastic processes as that of a Bernoulli trial as success and the remaining outcome(s)
as failure(s). In context of the present work, we define \textit{success} as the event in
which the lamb takes resort to the safe haven and \textit{failure} as the event in which
it is captured by the lion. This raises the following question: under what conditions can
the probability of success be maximized? An answer to this question is fixed in
the sense that given the values of motion parameters like the diffusion coefficients and
the initial locations of the lamb and the lion, we can estimate the probability of
success. However, if we introduce restarts in the system dynamics, then we can optimize
the probability of a successful completion of this Bernoulli trial \cite{belan2018restart}.

In the present work, we employ two restart protocols: one in which the rate of restart
is fixed, aka, Poissonian resetting \cite{evans2011diffusion,evans2018run,gupta2014fluctuating,
majumdar2015dynamical,singh2022general,ahmad2019first,ahmad2022first,masoliver2019telegraphic,
domazetoski2020stochastic,singh2021backbone,evans2014diffusion,pal2022inspection} and the other
in which the time between two restarts is fixed, aka sharp restarts \cite{pal2016diffusion,pal2017first,
chechkin2018random,eliazar2020mean,eliazar2021mean,eliazar2021tail,eliazar2022entropy,
eliazar2022diversity}. The reason for covering these two restart protocols is that they
lie at the two extremes of the class of renewal restart protocols: Poissonian resetting being
memoryless and sharp restart retaining its entire memory. Notwithstanding the extensive literature
addressing the effects
of both Poissonian and sharp restarts, studies addressing stochastic processes ending in
more than one ways have been rather limited \cite{belan2018restart,chechkin2018random,
pal2019first} and specific examples addressing the effect of sharp restarts on Bernoulli
trials is still missing, to the best of our knowledge. In order to pursue this goal, we
study the classic capture problem of a lamb being pursued by a lion in presence
of a safe haven for the lamb. Identifying the system as a couple of vicious Brownian
particles, we first provide the solution for the two particle problem and from thereon
estimate the survival probability for the lamb. Then we study the effect of Poisson
and sharp restarts on the Bernoulli trial estimating and comparing the exit probability
for success for the two restart protocols.

\textit{Two vicious random walkers with an absorbing wall:}
Consider a pair of vicious Brownian particles on the positive half line with $0 \le x_1
\le x_2 < \infty$ \cite{gabel2012can}.
The process ends when either the first walker reaches the haven at $x_1=0$ or when the
trajectories of the two particles cross each other, that is $x_1 = x_2$, at which point
the two vicious walkers kill each other. The Fokker-Planck equation (FPE) describing
the probability density function (PDF) of the process is
\begin{align}
\label{fpe}
\dd_t p = D_1 \dd^2_1 p + D_2 \dd^2_2 p,
\end{align}
where $\dd_t \equiv \fr{\dd}{\dd t}, ~\dd^2_i \equiv \fr{\dd^2}{\dd x^2_i}, ~i = 1, 2$
and $p \equiv p(x_1,x_2,t)$. The initial condition for the FPE in (\ref{fpe}) is $p(x_1,x_2,0) =
\delta(x_1-a_1)\delta(x_2-a_2)$ with $a_1 < a_2$ alongwith the boundary conditions
$p(x_1=0,x_2=x>0,t) = 0$ (the lamb reaching the haven) and $p(x_1=x,x_2=x,t) = 0$ (lion kills the
lamb). Without any loss of generality we assume that the two Brownian particles have
identical diffusion coefficients, that is, $D_1 = D_2 = D$. The FPE in Eq.~(\ref{fpe})
can be solved using the method of images and its solution can be written as an anti-symmetric
linear combination (see Fig.~{5} in Ref.~\cite{fisher1984walks}):
\begin{align}
\label{pxt}
&p(x_1,x_2,t) = f(x_1,x_2,t) - f(x_2,x_1,t) - f(-x_1,x_2,t)\nonumber\\
&+ f(x_2,-x_1,t)- f(-x_2,-x_1,t) + f(-x_1,-x_2,t)\nonumber\\
&- f(x_1,-x_2,t) + f(-x_2,x_1,t),
\end{align}
where $f(x_1,x_2,t) = \fr{1}{4\pi Dt}\exp\Big\{-\fr{(x_1-a_1)^2 + (x_2-a_2)^2}{4Dt}\Big\}$
is the PDF of a pair of non-interacting Brownian particles in one dimension. It is straightforward
to see that $p(x_1,x_2,t)$ satisfies the initial and boundary conditions complementing Eq.~(\ref{fpe}).
The asymmetric linear combination in (\ref{pxt}) is robust against the intrinsic details of
the random walk, be it the Brownian motion considered here or the random walk with discrete times of
Ref.~\cite{fisher1984walks}. We cannot overemphasize on the importance of this result.
 
\textit{First passage time distribution:}
The PDF in (\ref{pxt}) allows us to estimate the survival probability:
$q(t) = \int^\infty_0 dx_1 \int^\infty_{x_1} dx_2
~p(x_1,x_2,t)$ and from there the first passage time distribution (FPTD) reads
$F(t) = -\fr{d}{dt}q(t)$ leading to
\begin{align}
\label{fptd}
F(t) &= -D\Big[\int^\infty_0 dx_2 \fr{\dd p}{\dd x_1}\Big|^{x_2}_0
+  \int^\infty_0 dx_1 \fr{\dd p}{\dd x_2}\Big|^\infty_{x_1}\Big]\nonumber\\
&= \fr{e^{-1/8t}}{\sqrt{8\pi t^3}}\Big[\text{erf}\Big(\fr{3}{\sqrt{8t}}\Big)
+ \sqrt{2}\text{erf}\Big(\fr{1}{\sqrt{t}}\Big)e^{-1/8t}\nonumber\\
&- 2\sqrt{2}\text{erf}\Big(\fr{1}{2\sqrt{t}}\Big)e^{-7/8t}
- 3\text{erf}\Big(\fr{1}{\sqrt{8t}}\Big)e^{-1/t}\Big],
\end{align}
wherein we have chosen $a_1 = 1,~a_2 = 2$ and $D = 1$ to simplify the presentation.
The integral leading to Eq.~(\ref{fptd}) above has been  evaluated using MAXIMA.
Using the small argument approximation for the exponential and error function we
have $F(t) \stackrel{t \to \infty}{\sim} 1/\pi t^3$. As a result, $q(t)
\stackrel{t \to \infty}{\sim} 1/t^2$, previously derived using a wedge
domain in Ref.~\cite{gabel2012can}. This implies that the unconditional mean first passage
time is finite, and in addition, it is the only finite moment possessed by the FPTD
in Eq.~(\ref{fptd}). From this we can write the expressions for the conditional FPTDs,
the process terminating either in a success or failure. Define $F_1(t)$ as the
distribution of first passage times that the process ends when the first particle
reaches the origin irrespective of the location of the second particle, that is,
a successful completion of the Bernoulli process. Similarly, let $F_2(t)$
denote the conditional FPTD for the process to end by the two vicious walkers killing
each other, that is, a failure. Then, $F(t) = F_1(t) + F_2(t)$ and from Eq.~(\ref{fptd})
we obtain:
\begin{figure}
\includegraphics[width=0.5\textwidth]{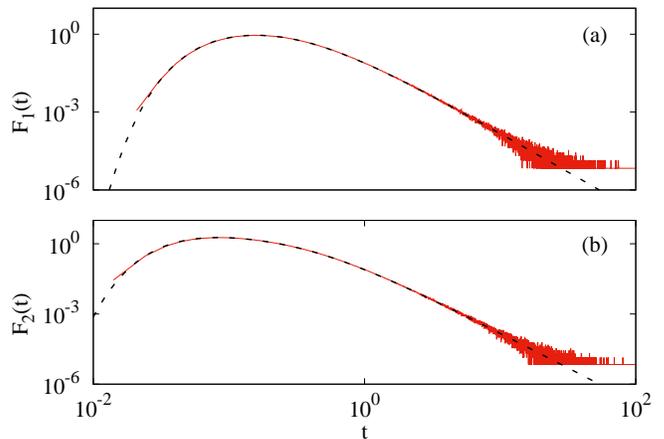}
\caption{Conditional FPTDs $F_1(t)$ and $F_2(t)$ for the pair of vicious Brownian particles.
The red line is the numerical estimate and the black dashed lines are the expressions
from Eq.~(\ref{cfpt}). Parameter values are $a_1 = 1, ~a_2 = 2, ~D = 1$. For these values:
$\mathcal{E}_1 \approx 0.41$.}
\label{fig1}
\end{figure}
\begin{subequations}
\label{cfpt}
\begin{align}
\label{cfpts}
F_1(t) &= \fr{e^{-1/4t}}{\sqrt{4\pi t^3}}\Big[\text{erf}\Big(\fr{1}{\sqrt{t}}\Big)
-2\text{erf}\Big(\fr{1}{\sqrt{4t}}\Big)e^{-3/4t}\Big],\\
F_2(t) &= \fr{e^{-1/8t}}{\sqrt{8\pi t^3}}\Big[\text{erf}\Big(\fr{3}{\sqrt{8t}}\Big)
-3\text{erf}\Big(\fr{1}{\sqrt{8t}}\Big)e^{-1/t}\Big].
\end{align}
\end{subequations}
We test our analytical results by numerically estimating the conditional FPTDs for
success and failure. We obtain these by numerically solving the Langevin equations
\begin{subequations}
\label{dyn}
\begin{align}
\dot{x}_1 &= \eta_1(t),\\
\dot{x}_2 &= \eta_2(t).
\end{align}
\end{subequations}
In (\ref{dyn}), $\eta_1(t)$ and $\eta_2(t)$ are two independent Gaussian random deviates
with mean zero and identical delta correlated variance, that is, $\langle \eta_i(t) \eta_i(t')
\rangle = 2D\delta(t-t')$ for $i = 1,2$ with $D = 1$. At $t = 0$ the two walkers are at
$x_1 = 1$ and $x_2 = 2$ and the process ends when either $x_1 = 0$ or $x_1 = x_2$.
Let $\mathcal{E}_1$ be the exit probability for the termination of the process by
the first particle reaching the origin and $\mathcal{E}_2$ for the trajectories crossing
each other. If $F_{1,n}(t)$ and $F_{2,n}(t)$ are the numerically estimated normalized
histograms for the conditional first passage times, then $F_i(t) = \mathcal{E}_i F_{i,n}(t)$.
We study this relation in Fig.~\ref{fig1} and find a good agreement between the analytical and
numerical estimates of the conditional FPTDs.

\textit{Restarting the Bernoulli process:}
For reasons discussed in Ref.~\cite{singh2022capture}, we reset the two vicious
walkers at the exact same moment. Furthermore, the time between two successive restarts is
chosen for the purpose of simplicity to be either an exponentially distributed random variable (Poissonian resetting) or a
fixed quantity (sharp resetting).
If $T$ is the time of unconditional completion of
the Bernoulli trial under consideration and $R$ is the time of restart of the process,
then the probability of success is:
$p = \fr{\langle I(T<R) y_T \rangle}{\langle I(T<R) \rangle}$ \cite{belan2018restart},
where $y_T$ is an auxiliary random variable taking value one with probability
$F_1(t)/F(t)$. Now, if $\langle T^s \rangle$ is the mean completion time of a successful trial
then $\langle T^s \rangle = \fr{\langle I(T>R) R \rangle}{\langle I(T<R) \rangle}
+ \fr{\langle I(T<R) y_T T \rangle}{\langle I(T<R)y_T \rangle}$ \cite{belan2018restart}.
Let us now discuss Poisson and sharp restart protocols one by one.

\begin{figure}
\includegraphics[width=0.5\textwidth]{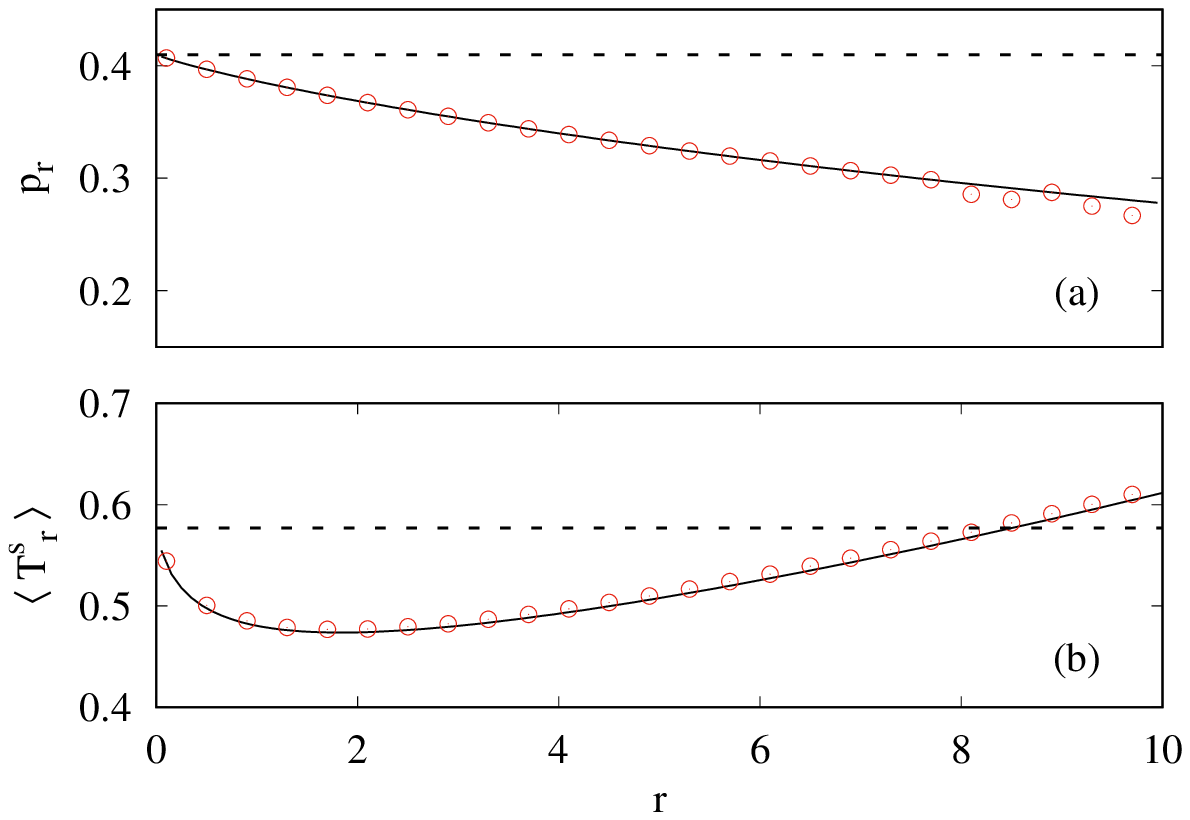}
\caption{Effect of Poisson restarts: (a) Numerically estimated probability of success
$p_r$ (red circles) vs
the analytical result from Eq.~(\ref{pr}) (black solid line). The black dashed line
represents $\lim_{r \to 0} p_r$, success probability in absence of any restarts.
(b) Numerically estimated mean time for a successful completion $\langle T^s_r \rangle$
(red circles) vs the analytical result from Eq.~{\ref{tsr}} (black solid line). Black
dashed line represents $\lim_{r \to 0}\langle T^s_r \rangle$. Parameter values are
$a_1 = 1, ~a_2 = 2, ~D = 1$.}
\label{fig3}
\end{figure}
For Poissonian resetting at a rate $r$, the PDF of restart times is $P^r(R) = re^{-rR}$.
As a result, the probability of success and the mean time for a successful
completion respectively read \cite{belan2018restart}:
\begin{subequations}
\label{rate}
\begin{align}
\label{pr}
p_r &= \fr{\tl{F_1}(r)}{\tl{F}(r)},\\
\label{tsr}
\langle T^s_r \rangle &= \langle T_r \rangle - \fr{d}{dr}\ln p_r,
\end{align}
\end{subequations}
where $\langle T_r \rangle = \fr{1-\tl{F}(r)}{r\tl{F}(r)}$ \cite{reuveni2016optimal}
denotes the mean time of completion of the Bernoulli trial, either with a success
or as a failure; and $\tl{F}(r) = \int^\infty_0 dt~e^{-rt}F(t)$ is the Laplace transform
of $F(t)$. We compare these results for the capture problem against their numerical
solution by solving Eq.~(\ref{dyn}) under Poissonian restarts at a rate $r$ and
find good agreement between the two (see Fig.~\ref{fig3}). It is evident from Fig.~
\ref{fig3}(a) that $p_r$ is a monotonically decreasing function of the restart rate
$r$, while the mean time for successful completion $\langle T^s_r \rangle$ exhibits
a minima, as seen from Fig.~\ref{fig3}(b). This implies that while resetting makes it
slightly less probable for the lamb to make it to the
safe haven, the time to reach the
safe haven can be minimal, for example, for $r \approx 2$.
For higher values of restart rate like $r \approx 10$, the
lamb is walking a slippery
slope where it takes a longer time to reach the haven 
and the chances of it doing so are also
severely diminished, thanks to the fact that it keeps returning home.
It should be noted at this point that the Laplace
transforms in Eq.~(\ref{rate}) have been evaluated via numerical integration 
\cite{press1986numerical}.
Let us now move on to studying the Bernoulli trial under sharp resetting.

\begin{figure}
\includegraphics[width=0.5\textwidth]{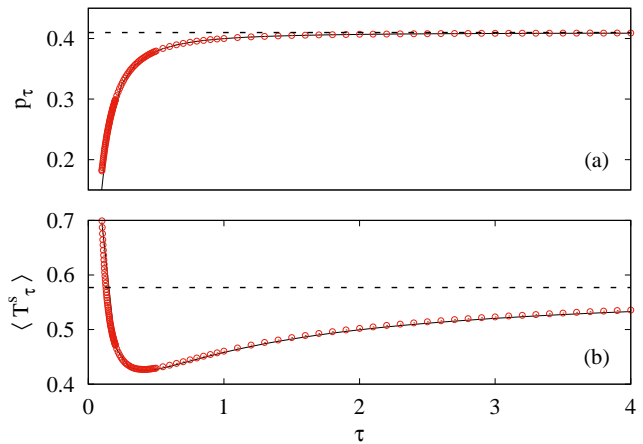}
\caption{Effect of sharp restarts: (a) Numerically estimated probability of success
$p_\tau$ (red circles) vs
the analytical result from Eq.~(\ref{p_t}) (black solid line). The black dashed line
represents $\lim_{\tau \to \infty} p_\tau$, success probability in absence of any restarts.
(b) Numerically estimated mean time for a successful completion $\langle T^s_\tau \rangle$
(red circles) vs the analytical result from Eq.~{\ref{mfpt_ts}} (black solid line). Black
dashed line represents $\lim_{\tau \to \infty}\langle T^s_\tau \rangle$. Parameter values are
$a_1 = 1, ~a_2 = 2, ~D = 1$.}
\label{fig2}
\end{figure}
For sharp resetting the PDF of restart times is $P^\tau(R) = \delta(R-\tau)$, where
$\tau$ is the time of sharp restart. Then $\langle I(T<R) y_T\rangle = \int^\infty_0
dR~P^\tau(R)\int^\infty_0 dt~F(t)I(T<R)y_T = \int^\infty_0 dt~F_1(t) \int^\infty_t
dR~\delta(R-\tau) = \int^\tau_0 dt~F_1(t)$ where we have reversed the order of integration
in the second equality and the delta function term contributes only when $t \le \tau$.
In a similar manner we have $\langle I(T<R) \rangle = \int^\tau_0 dt~F(t)$, from where
follows the probability of success under sharp resetting
\begin{align}
\label{p_t}
p_\tau = \fr{\int^\tau_0 dt ~F_1(t)}{\int^\tau_0 dt ~F(t)}.
\end{align}
Evaluating the remaining integrals we get $\langle I(T>R) R \rangle = \tau \int^\infty_\tau
dt~F(t)$ and $\langle I(T<R) y_T T \rangle = \int^\tau_0 dt~t F_1(t)$ leading to
the mean time for a successful completion: $\langle T^s_\tau \rangle = \fr{\tau
\int^\infty_\tau dt~F(t)}{\int^\tau_0 dt~F(t)} + \fr{\int^\tau_0 dt~t F_1(t)}{\int^\tau_0
dt~F_1(t)}$ which can be re-written as
\begin{align}
\label{mfpt_ts}
\langle T^s_\tau \rangle = \langle T_\tau \rangle
- (1-p_\tau)\fr{\int^\tau_0 dt~t F_2(t)}{\int^\tau_0 dt~F_2(t)}.
\end{align}
In the above equation, $\langle T_\tau \rangle = \fr{\int^\tau_0 dt~t F(t)}{\int^\tau_0 dt~F(t)}
+ \fr{\tau\int^\infty_\tau dt~F(t)}{\int^\tau_0 dt~F(t)}$ is the mean time of completion
of the Bernoulli trial in presence of sharp resetting and $F_2(t)$ is defined in
Eq.~(\ref{cfpt}). It is interesting to see the close analogy between
the equations for Poisson and sharp restarts. We now compare the analytical results
of Eq.~(\ref{p_t}) and (\ref{mfpt_ts})
with numerical solution of the Langevin equations (\ref{dyn}) under sharp resetting
and find excellent agreement between the two approaches in Fig.~\ref{fig2}. Furthermore,
the success probability under sharp restarts $p_\tau$ asymptotically approaches its
value in absence of any restarts in a monotonic way and remains less than
$\lim_{\tau \to \infty} p_\tau$ (see Fig.~\ref{fig2}(a)). On the other hand, the mean
time taken by the lamb to successfully
reach the safe haven $\langle T^s_\tau \rangle$ exhibits a non-monotonic
dependence (in sharp contrast with $p_\tau$)
on the restart time $\tau$. This
implies that sharp resetting is advantageous for the
lamb as it is able to quickly take resort to the safe haven
as compared to the case when there are no restarts.
Unlike its Poissonian counterpart, a
sharp restart of the Bernoulli trial with high value of $\tau$ is advantageous for
the lamb, as its probability to make it to the safe haven is close to $\mathcal{E}_1$,
and this mode of completion takes a lesser amount of time on average. This prompts
us to make an explicit comparison between the two restart protocols, and more so
their relation to the dynamics of the Bernoulli trial without restarts.
We proceed with this goal in the next section.

\textit{Comparing Poissonian restart with sharp restart:}
\begin{figure}
\includegraphics[width=0.5\textwidth]{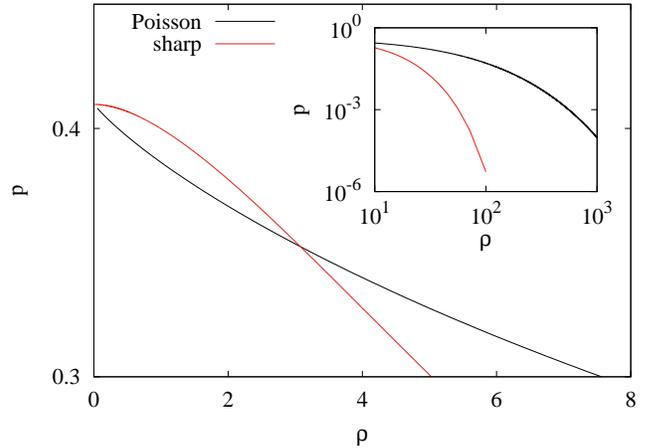}
\caption{Comparing success probability for Poisson and sharp restarts. Inset shows
the behavior at very large rates. Here the rate $\rho$ denotes either $r$ or $1/\tau$ depending
on the restart protocol under consideration, viz. Poisson or sharp, respectively.
Parameter values are $a_1 = 1, ~a_2 = 2, ~D = 1$.}
\label{fig4}
\end{figure}
In order to compare Poissonian and sharp restart for the Bernoulli trial under consideration, let
us define $1/\tau$ as the rate of sharp restart, where $\tau$ is the time of sharp restart.
This definition puts the two restart protocols on same footing and we define $\rho$ as the
rate of restart, with $\rho = r$ for Poisson restart and $\rho = 1/\tau$ for sharp restart.
As we have seen above, for both Poisson and sharp restart protocols, we have $\lim_{r \to 0}p_r
= \lim_{\tau \to \infty}
{p_\tau} = \mathcal{E}_1$. This is also seen in Fig.~\ref{fig4} for $\rho$ in the
neighborhood of zero. However, the approach of $p_r$ and $p_\tau$ to $\mathcal{E}_1$ is completely
different in that the second derivative $d^2 p/d\rho^2$ near $\rho \approx 0$ is
positive or negative depending on
whether we consider Poisson or sharp restart protocol (see Fig.~\ref{fig4}). This can
be understood as follows. For sharp restart $\rho$ near zero means that the time interval
between two successive restarts $\tau$ is very large, which means that the Bernoulli trial
stops without being practically perturbed by any restart event (see the near horizontal behavior
of $p$ for small $\rho$ in Fig.~\ref{fig4}). On the other hand, since $\langle T \rangle <
\langle T^s \rangle$ (from Eqs.~(\ref{fptd}) and (\ref{cfpts})), we have for Poissonian restarts:
$p_r \stackrel{r \to 0}{\sim} \mathcal{E}_1 + \mathcal{E}_1(\langle T \rangle - \langle T^s \rangle)r$,
which leads to $dp_r / dr < 0$ for $r \to 0$. Note that we can take this analysis no further as
the higher order moments of the FPTD $F(t)$ do not exist. The difference in the signs of the
second derivative of success probability eventually lead to $p_\tau < p_r$ for large $\rho$
with $\lim_{\rho \to \infty} p = 0$ (see inset in Fig.~\ref{fig4}). This follows simply from
the fact that for large $\rho$ both the vicious walkers are reset to their initial locations
very rapidly, making it practically impossible for the lamb to make it to the safe haven.
However, the fact that lion is also going to its den every now and then, makes the mean
time taken by the lamb to reach the safe have
extremely large (see Figs.~\ref{fig3} and {\ref{fig2}} for a comparison).

\begin{figure}
\includegraphics[width=0.5\textwidth]{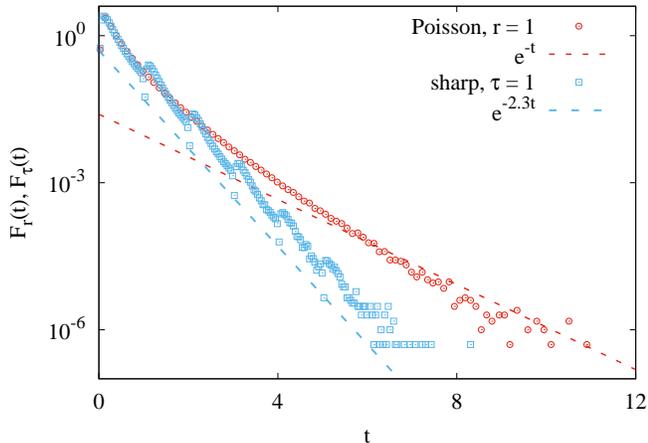}
\caption{Comparing the FPTDs $F_r(t)$ and $F_\tau(t)$ under Poissonian (red)
and sharp (blue) restarts respectively. The symbols are numerically estimated and the
dashed lines represent the analytical approximations obtained via inverting the Laplace
representations. The periodic behavior of $F_\tau(t)$ is evident. Parameter values
are $a_1 = 1, ~a_2 = 2, ~D = 1$.}
\label{fig5}
\end{figure}
Let us now discuss the effect of resetting on the FPTD $F(t)$. Under Poissonian resetting
at a rate $r$, the FPTD is known to be: $\tl{F}_r(s) = \fr{\tl{F}(s+r)}{\fr{s}{s+r} +
\fr{r}{s+r}\tl{F}(s+r)}$ \cite{reuveni2016optimal} and from here follows the FPTD under
Poissonian resetting via the Bromwich integral \cite{arfken1999mathematical}:
\begin{align}
\label{fr_fptd}
F_r(t) = \fr{1}{2\pi i}\int^{\gamma + i\infty}_{\gamma - i\infty}
ds~\fr{(s+r)\tl{F}(s+r)}{s + r \tl{F}(s+r)} e^{st}.
\end{align}
While an exact evaluation of the above integral is difficult, we can obtain the long-time
behavior of $F_r(t)$ by looking at the pole of $\tl{F}_r(s)$ closest to the origin. If $s_{0,r}$
is the pole nearest to zero, then it solves the equation: $0 = s + r \mathcal{L} [e^{-rt} F(t)]$,
where $\mathcal{L}$ denotes Laplace transform. We evaluate the Laplace transform for
a specific value of $r = 1$, and by numerically estimating $\mathcal{L} [e^{-rt} F(t)]$
we find $s_{0,r} \approx -1.0$. As a result, for $t \gg 1$ we have $F_r(t) \approx e^{-s_{0,r}t}$.
We compare this approximate result in Fig.~\ref{fig5}, wherein we estimate $F_r(t)$ by
numerically solving the Langevin equations in (\ref{dyn}) under Poissonian resetting with
$r = 1$. It is evident from the figure that we have a reasonable agreement between 
the analytical and numerical estimates, though we would not call it good. The reason for
this difference is that numerical estimations suggest $F_r(t) \approx e^{-1.25 t}$, which
is an error of about $20\%$ when compared to the analytical approximation
$F_r(t) \approx e^{-t}$. With an exact representation of the Laplace transform $\tl{F}(s)$
unavailable, we cannot provide a plausible explanation for this difference here.

For sharp resetting, the FPTD in Laplace domain reads \cite{pal2017first,singh2022capture}:
\begin{align}
\label{ft_fptd}
\tl{F}_\tau(s) = \fr{\int^\tau_0 dt~F(t)e^{-st}}{1-e^{-s\tau}\int^\infty_\tau dt~F(t)},
\end{align}
and its Laplace inversion at long times
is determined by the pole $s_{0,\tau}$ of $\tl{F}_\tau(s)$ closest to zero. It
is given by the solution of the equation: $0 = 1-e^{-s_{0,\tau}\tau}\int^\infty_\tau dt~F(t)
\Rightarrow s_{0,\tau} = \fr{1}{\tau}\log\Big(\int^\infty_\tau dt~F(t)\Big)$. For a specific value
like $\tau = 1$ results in $s_{0,\tau} \approx -2.3$ which, at large times
leads to $F_\tau(t) \approx e^{-s_{0,\tau} t}$.
Let us now look at the properties of $\tl{F}_\tau(s)$ in some more detail. The defining
property of sharp resetting is that it introduces a periodicity in the system, with the
period being $\tau$, the time interval between two sharp restarts. Furthermore, if we consider
a periodic function with a period $\tau$, that is, $h(t+\tau) = h(t)$, then its Laplace
transform reads \cite{spiegel1965laplace,schiff1999laplace}:
\begin{align}
\label{hts}
\tl{h}(s) = \fr{\int^\tau_0 dt~h(t)e^{-st}}{1-e^{-s\tau}}.
\end{align}
A quick look at Eqs.~(\ref{ft_fptd}) and (\ref{hts}) shows the degree of their similarity,
except for the appearance of the term $\int^\infty_\tau dt~F(t)$ in the denominator of the
fraction defining $\tl{F}_\tau(s)$ in Eq.~(\ref{ft_fptd}). Now, with the information that
$F_\tau(t) \approx e^{-s_{0,\tau}t}$ for large $t$, we can discern that the FPTD $F_\tau(t)$
has a periodic structure (with period $\tau$) and its envelope decaying exponentially.
It is to be noted at this point that this behavior of the FPTD is generic to any first
passage process under sharp resetting, and not limited to the Bernoulli trial under
consideration. For the Bernoulli trial, however, we can make a comparison of our analytical
approximation against numerical calculations. We see from Fig.~\ref{fig5} that $F_\tau(t)$
does exhibit a periodic behavior with an envelope tracing the curve $F_\tau(t) \approx
e^{-2.3t}$ (obtained via Laplace inversion). Furthermore, the period of the FPTD $F_\tau(t)$
is $\tau = 1$, as explained above.

It is to be noted that we have studied the Bernoulli trial problem under restarts for 
parameter values $(a_1, a_2, D)
= (1,2,1)$, though we checked for other parameter values like $(a_1, a_2, D) = (1/2, 2, 1)$
and found similar behavior. It is for this reason that we report only the former set of parameters.

\textit{Conclusions:} If we ask ourselves one question, what is the quintessential problem
in life, we shall almost always come to one answer: the problem is choice. Motivated by this
line of thought, we study the capture problem wherein a hungry lion pursues a lamb in presence
of a safe haven for the lamb. This is one of the classic problems in nonequilibrium statistical
physics, and presents us with an example of a first passage process which can terminate in
two ways: either the lamb reaches the safe haven (success) or is killed by the lion (failure).
Following the seminal work of Fisher \cite{fisher1984walks}, we are able to obtain the
exact solution for the two-particle problem. We find that the distribution of first passage times
possesses a finite mean, though no higher moments exist. We further study this problem
under Poissonian and sharp restarts, and find that the probability of success exhibits
a monotonic dependence on the rate of restart for Poissonian restart, or the time of
sharp restart. In addition, the FPTDs under restarts exhibit exponentially decaying tails,
and we find a reasonable agreement between the analytical approximations and the
numerical estimates.

The importance of capture problems, and a limited amount of literature studying Bernoulli
trials under restarts makes this study a timely work. While we have chosen two identical
Brownian particles, going to a set of non-identical but vicious walkers (different diffusion
coefficients $D_1$ and $D_2$) is straightforward. However, the more interesting cases wherein
the walkers are either a pair of vicious run and tumble particles \cite{le2019noncrossing}
or a run and tumble particle viciously interacting with another Brownian particle are worth
considering.

\textit{Acknowledgments}: RKS thanks the Israel Academy of Sciences and Humanities (IASH)
and the Council of Higher Education (CHE) Fellowship. TS acknowledges financial support
by the German Science Foundation (DFG, Grant number ME~1535/12-1). TS is supported by the
Alliance of International Science Organizations (Project No.~ANSO-CR-PP-2022-05). TS is
also supported by the Alexander von Humboldt Foundation. SS thanks Kreitman Fellowship and HPC
facility at Ben-Gurion University.

\bibliography{ref.bib}

\end{document}